%% file: ms_astroph.tex
\documentclass[12pt,preprint]{aastex}

\usepackage{ulem}

\slugcomment{Accepted to ApJ}

\shorttitle{Supernovae triggered Star Formation}
\shortauthors{Chikako Yasui et al.}

\begin{document}


\title{DEEP NEAR-INFRARED IMAGING OF \\
AN EMBEDDED CLUSTER \\ 
IN THE EXTREME OUTER GALAXY:\\
CENSUS OF SUPERNOVAE TRIGGERED \\ STAR FORMATION
\footnote{Based on data collected at Subaru Telescope,
which is operated by the National Astronomical Observatory of Japan.}}



\author{Chikako Yasui and Naoto Kobayashi}
\affil{Institute of Astronomy, University of Tokyo, 2-21-1 Osawa,
Mitaka, Tokyo 181-0015, Japan}
\email{ck$_-$yasui@ioa.s.u-tokyo.ac.jp}

\author{Alan T. Tokunaga}
\affil{Institute for Astronomy, University of Hawaii, 2680 Woodlawn
Drive, Honolulu, HI 96822, USA}

\author{Hiroshi Terada}
\affil{Subaru Telescope, National Astronomical Observatory of Japan, 84 Pukihae Street, Hilo, HI 96720, USA}

\and

\author{Masao Saito} 
\affil{ALMA Project, National Astronomical Observatory of Japan, 2-21-1
Osawa, Mitaka, Tokyo 181-8588, Japan}




\begin{abstract}
While conducting a near-infrared (NIR) survey of ``Digel Clouds'', which
are thought to be located in the extreme outer Galaxy (EOG), Kobayashi
\& Tokunaga found star formation activity in ``Cloud 2'', a giant
molecular cloud at the Galactic radius of $\sim$ 20 kpc.  Additional
infrared imaging showed two embedded young clusters at the densest
regions of the molecular cloud.
Because the molecular cloud is located in the vicinity of a supernova
remnant (SNR) HI shell, GSH 138-01-94, it was suggested that the star
formation activity in Cloud 2 was triggered by this expanding HI shell.  We obtained deep J (1.25$\,\mu$m), H
(1.65$\,\mu$m) and K (2.2$\,\mu$m) images of one of the embedded
clusters in Cloud 2 with high spatial resolution
(FWHM $\sim 0\farcs 3$) 
and high sensitivity (K $\sim$ 20 mag, 10$\sigma$).
We identified 52 cluster members.
The estimated stellar density ($\sim$ 10 pc$^{-2}$) suggests that the 
cluster is a T-association.
This is the deepest NIR imaging of an embedded
cluster in the EOG.  The observed K-band luminosity function (KLF)
suggests that the underlying initial mass function (IMF) of the cluster
down to the detection limit of $\sim 0.1 M_\odot$ is not significantly
different from the typical IMFs in the field and in the near-by star
clusters.  The overall characteristics of this cluster appears to be
similar to those of other embedded clusters in the far outer Galaxy.  
%
%
The estimated age 
of the cluster from the KLF
, which is less than 1 Myr, is consistent with the view that the star
formation was triggered by the HI shell whose age was estimated at 4.3
Myr (Stil \& Irwin).  The 3-dimensional geometry of SNR shell, molecular
cloud and the embedded cluster, which is inferred from our data, as well
as the cluster age strongly suggest that the star formation in Cloud 2
was triggered by the SNR shell.

\end{abstract}





  
\section{Introduction} \label{sec:INTRO}

The extreme outer Galaxy (EOG), whose Galactic radius ($R_g$) is more
than 18 kpc, is known to be a very low-density and low-metallicity
region.  Because the distribution limit of population I and II stars are
18-19 kpc and 14 kpc, respectively, the star forming process in such
environment has not been studied
as well as near-by star forming regions or star forming regions 
in the inner Galaxy 
due to the faintness of the sources.
The EOG also serves as an excellent laboratory for the
study of star forming process because there is no complex star formation
history as in the inner Galaxy.

``Cloud 2'' is one of the molecular clouds in the outermost Galaxy
region discovered by \cite{Digel1994}
based on CO observations of distant HI peaks in the Maryland-Green
Bank survey \citep{WW1982}. 
It is located at a very
large Galactic radius ($R_g$ = 18$-$28 kpc) 
%
in the second quadrant of the Galaxy ($l \sim 130^\circ$).
It is the largest molecular cloud (M $\sim$ 1 $\times$ 10$^4$
M$_\odot$; \citet{Digel1994}) among those in the outermost region.
The Galactic radius of Cloud 2 has been 
estimated by various methods:
$R_g = 28$ kpc (heliocentric distance: $D=$ 21 kpc)
from the kinematic distance of CO \citep{Digel1994},
$R_g =$ 23.6 kpc ($D = 16.6$ kpc)
from the latest HI observation \citep{Stil2001},
and $R_g \sim$ 15$-$19 kpc ($D=$ 8.2$-$12 kpc)
from the optical 
spectroscopy
 of a B-type star, 
MR-1 \citep{Muzzio1974, Smartt1996},
which is thought to be associated with Cloud 2 \citep{de Geus}.
Throughout this paper we adopt $R_g$=19 kpc ($D=$12 kpc)
because it is about the mean value of all the estimated distances 
and because spectroscopic distance of stars should be more accurate than
kinematic distance.
Metallicity at the Galactic radius of Cloud 2 is estimated at 
$\sim$ -1.0 dex from the standard metallicity curve by
\citet{Smartt1997}.
In fact, the metallicity of the B-type star, MR-1, is measured at
$-$0.5 to $-$0.8 dex 
\citep{Smartt1996, Rolleston2000}
and that of the Cloud 2 itself, is measured at $\sim$ -0.7 dex from 
the radio molecular 
emission 
lines \citep{Lubowich}.
These metallicity values are comparable to that of LMC ($\sim -0.5$ dex)
and SMC ($\sim -0.9$ dex) \citep{Arnault}.
Also, Cloud 2 was found to be located in the vicinity of a large
SNR HI shell (GSH 138-01-94) which has an almost complete spherical
shape with a radius of 180 pc \citep{Stil2001}.

\citet{KT2000}
found young stellar objects in Cloud 2
with a wide field near-infrared survey.
\citet{NK2006}
 also found two young stellar clusters at the
northern and southern  CO peaks of Cloud 2 with deeper near-infrared
imaging with UH 2.2m 
telescope.
They suggested from the geometry of the clusters and SNR HI shell
that the star formation activity in Cloud 2 was triggered by the SNR HI
shell. 
Cloud 2 may be the clearest example of SN triggered star formation,
which makes use of the advantage of the ``star formation laboratory''
in  the EOG.
As a next step, 
we obtained deep high-resolution
images of the stellar clusters in Cloud 2 with the Subaru
8.2 m telescope for studying
the details of the SN triggered star formation.
Here we present the results for the stellar cluster in the northern 
CO peak of Cloud 2 (hereafter ``Cloud2-N cluster'').

\section{Observations and Data Reduction} \label{OBSandDATA}

\subsection{Subaru IRCS imaging} \label{IMAGE}
We have obtained J(1.25\,$\mu$m), H(1.65\,$\mu$m) and
K(2.2\,$\mu$m)-band deep images of the 
Cloud2-N cluster.
The observation was conducted on 2000 December 1 UT
with the Subaru Infrared Camera 
and Spectrograph IRCS (Tokunaga et al. 1998, Kobayashi et al. 2000,
Terada et al. 2004) with the pixel scale of 0$\farcs$058/pixel. The
entire infrared cluster was sufficiently covered with the $\sim$1$'$
field-of-view. IRCS employs the Mauna Kea Observatory 
(MKO) 
near-infrared
photometric filters 
(Tokunaga, Simons \& Vacca 2002). The total integration time was 600,
675, and 1350 sec for J, H, and K-bands, respectively. The observing
condition was photometric and the seeing was excellent
($\sim$0$\farcs$3) through the observing period.

\subsection{Data Reduction} \label{DATA}
All the data for each band were reduced with
IRAF\footnote{IRAF is distributed by the National Optical Astronomy
Observatories, which are operated by the Association of Universities
for Research in Astronomy, Inc., under cooperative agreement with the
National Science Foundation.} 
with standard procedures:
dark subtraction, flat-fielding, bad-pixel correction, median-sky 
subtraction, image shifts with dithering offsets, and combining.
The stellar FWHM 
in final images of $J, H, K$-bands are $0\farcs35, 0\farcs3,
0\farcs35$, respectively.
$JHK$ photometry has been performed using IRAF APPHOT
with a $0\farcs58$ diameter aperture (10 pixel) except for a binary with
$0\farcs27$ separation for which a $0\farcs17$ diameter aperture was
applied  with an aperture correction.
2MASS 02484460+5828284 
($J=15.44, H=14.58, K_S=14.38$)
in the field
was used as a photometric standard star 
after converting the 2MASS system to the MKO system 
using the color transformations from S. K. Legett (private comm.).
%
%
%
%
%
%
The flux uncertainty $\sigma$ in the $0\farcs58$ aperture was
estimated from the standard deviation of the flux in 900 independent
apertures in the blank sky.
Compared to the flux uncertainty estimated by APPHOT from pixel-to-pixel
variation, $\sigma$ is larger by a factor of 1.20, 1.14, 1.34 for $J, H,
K$-bands, respectively.
The resultant limiting magnitudes in $J, H, K$-bands were estimated at
21.12, 19.96, 20.03 mag (10$\sigma$),
respectively. In total 138 stars were detected in the $K$-band brighter
than this limiting magnitude.

\section{RESULTS} \label{sec:RESULTS}

\subsection{Identification of Cloud2-N Members} \label{sec:IDENT}
\begin{figure}
\epsscale{.80}
\includegraphics[scale=0.8]{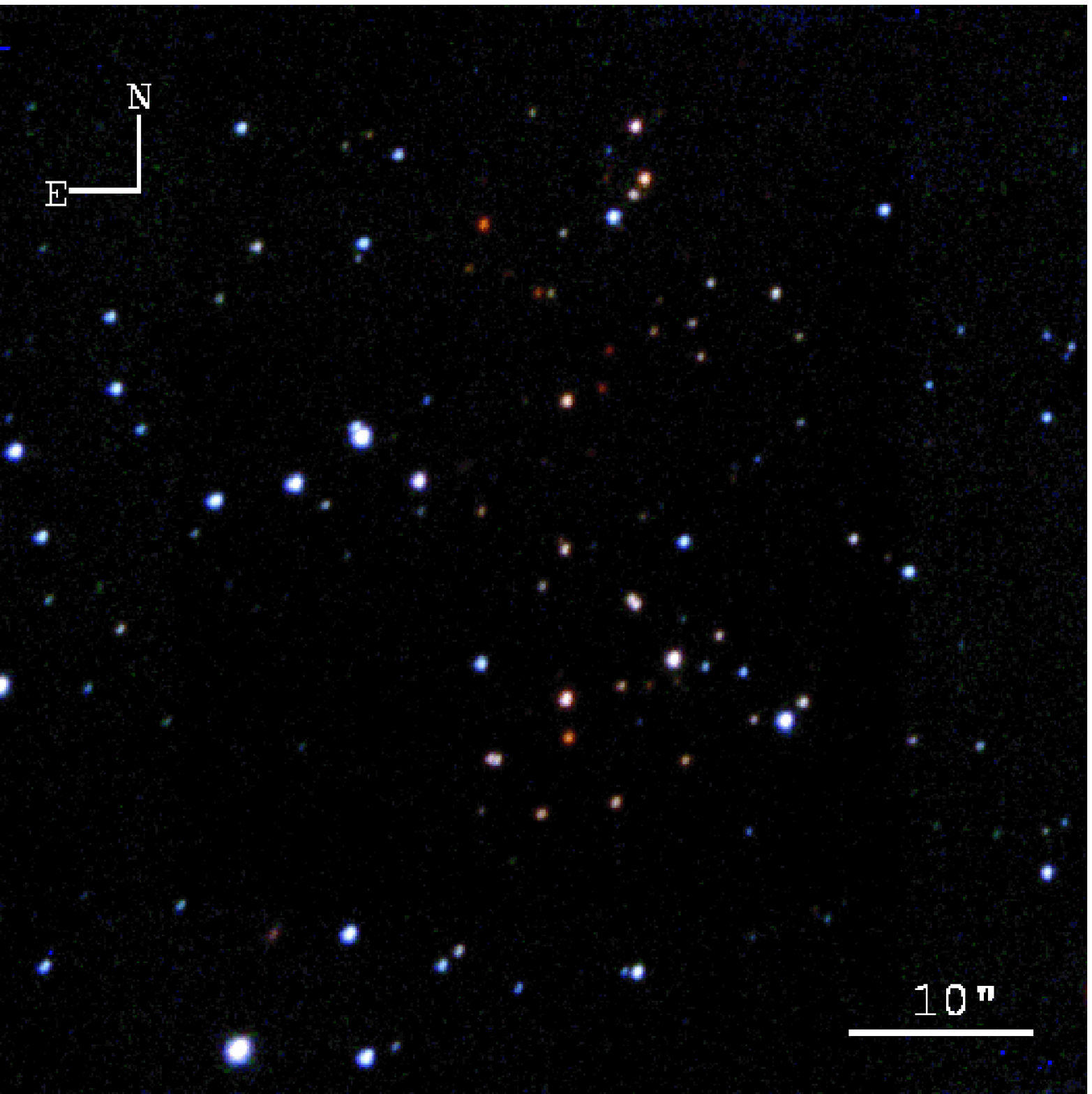}
\caption{
JHK pseudo color image of the Cloud2-N cluster.
North is up and east is left.
The the field of view is about $1\arcmin \times 1 \arcmin$.
The 
coordinate
of the field is 
$(\alpha, \delta) = (02^h48^m40^s, +58^d28^m56^s)$
with an uncertainty of a few arcsec.
}
\label{fig:3color}
\end{figure}

\begin{figure}
\includegraphics[angle=-90,scale=.60]{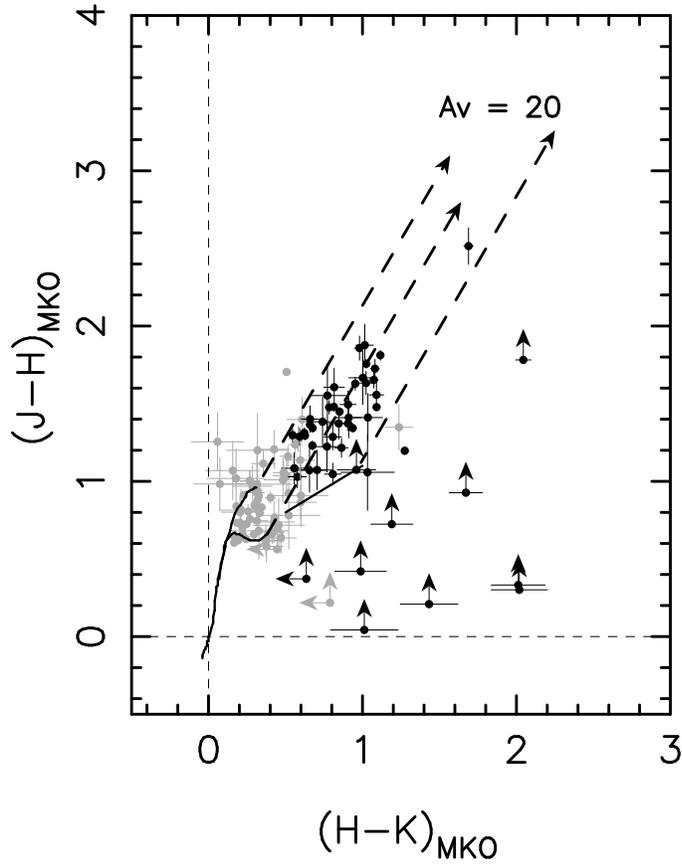}
\caption{(J-H) vs. (H-K) color-color diagram of the detected stars
in the IRCS field.
Stars detected with more than 5$\sigma$ in J, H-band and 10$\sigma$ in
 K-band are shown. 
Identified Cloud2-N cluster members and field stars are shown with black
and grey dots, respectively.
The dwarf and giant star tracks from 
\citet{Bessel}
and the classical T Tauri star locus from \citet{T Tauri}
are shown with thin lines and thick line respectively.
The reddening vectors \citep{RL} for A$_{\rm v}$ = 20 mag are shown
with dashed line. 
\label{fig:colcol}}
\end{figure}

We constructed a pseudo color picture of the observed field
 (Fig.~\ref{fig:3color}) and the
$J-H$ versus $H-K$
 color-color
diagram of all the detected sources 
(Fig.~\ref{fig:colcol}).
The cluster was first recognized by 
\citet{NK2006} 
as a loosely
packed cluster of red sources in
a $26'' \times 40''$ region
 (see Fig.~\ref{fig:3color}).
The colors of all the visually selected red sources from 
Fig.~\ref{fig:3color}
were found to have $H-K \ge 0.5$.
The sources with bluer color ($H-K < 0.5$) were found to be distributed
uniformly in the field. 
Therefore, we identified the cluster members with the following two
criteria:
1) $H-K \ge 0.5$, 2) distributed in $26'' \times 40''$ region at
the center.
As a result, we found 52 cluster members out of 138 sources in the
field. 
Because there are no foreground molecular clouds in front of Cloud 2
\citep{KT2000}, the red colors of the cluster members should originate
from the extinction by Cloud 2 itself 
as well as by circumstellar material 
(see section \ref{SNTSF} for more discussion). 
Contamination from the background sources should be very small in view
of the large $R_g$ of Cloud 2.

\subsection{Extinction and Disk Color Excess} \label{sec:AvDISK}
\begin{figure}
\includegraphics[angle=-90,scale=.50]{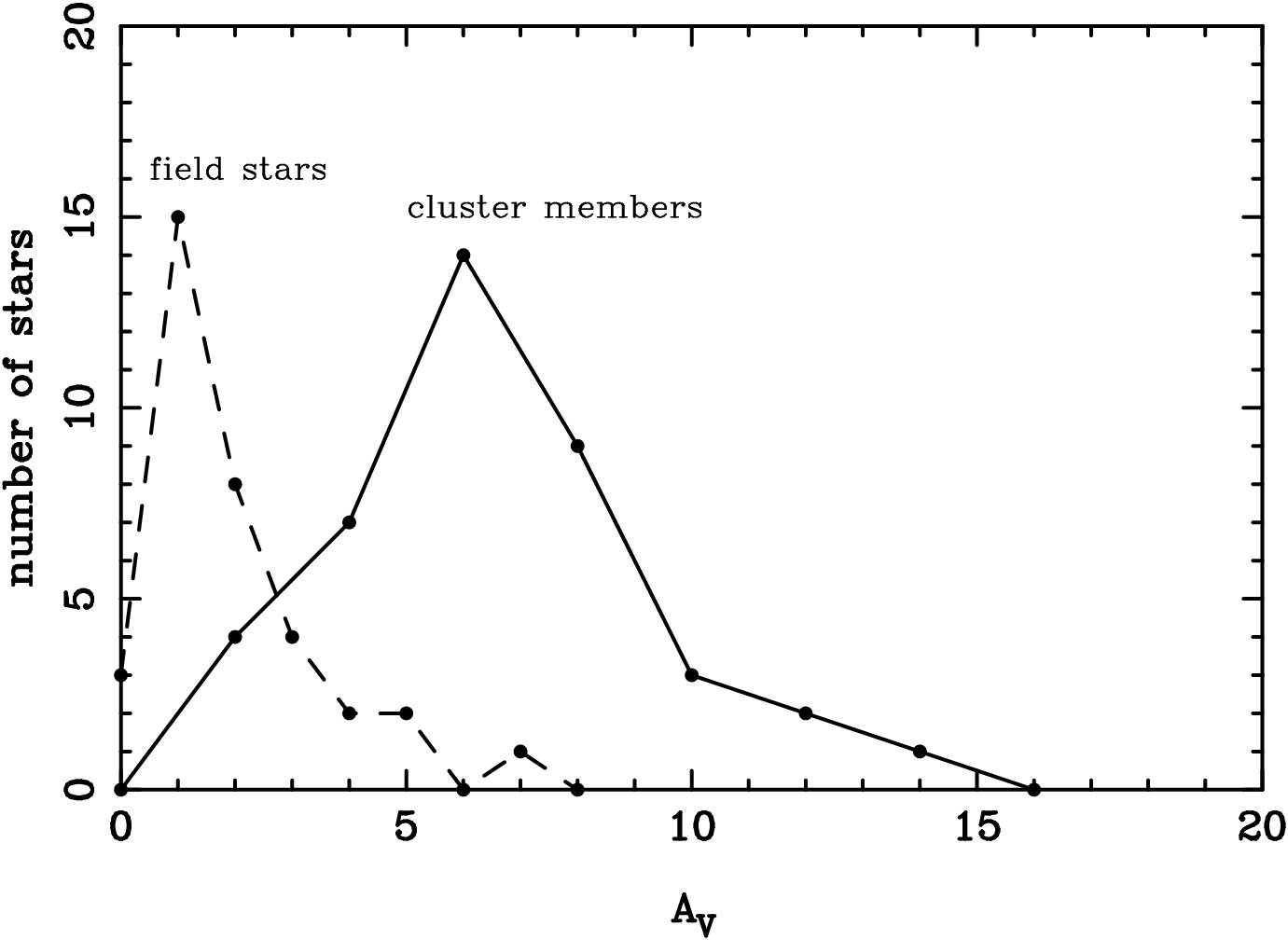}
\caption{Distributions of A$_{\rm v}$ for cluster members (solid line) and
field stars (dashed line).}
\label{fig:Av}
\end{figure}

\begin{figure}
\includegraphics[angle=-90,scale=.50]{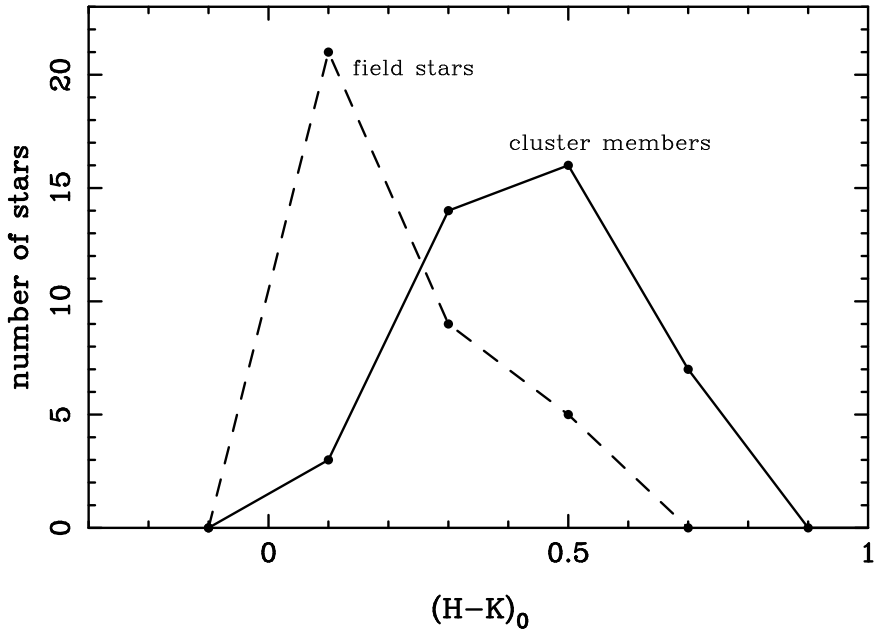}
\caption{Distributions of 
$(H-K)_0$
 for cluster members
 (solid line) and field stars (dashed line).}
\label{fig:HK}
\end{figure}

The extinction and disk color excess for each star were estimated using
the color-color diagram.
Those parameters will be used for constructing model KLFs in
section~\ref{sec:modelKLF}.
For reliable estimation of the parameters, 
stars only in the positions reddened from 
the classical T Tauri star (CTTS) locus \citep{T Tauri} 
and the dwarf locus (42 cluster members and 32 field stars)
are used.
For convenience the dwarf locus was approximated 
by the extension of the
CTTS locus drawn out to $H-K \sim 0.1$ mag.
In the color-color diagram  the extinction Av of each star was estimated
from the distance along the reddening vector \citep{RL}
between its location and the stellar loci.
The resultant Av distributions of the cluster members and the field
stars (Fig.~\ref{fig:Av}) have a peak at 6.4 mag (A$_{\rm K}$ = 0.72 mag) 
and 1.8 mag (A$_{\rm K}$ = 0.20 mag), respectively.
The distribution of the cluster members shows that the whole cluster is
reddened uniformly by a screen of dust with Av=6 mag.
This verifies
the selection method of cloud members described in
section~\ref{sec:IDENT}.
The distributions of the 
residual  color $(H-K)_0$ of cloud members and field stars
show a clear difference 
(Fig.~\ref{fig:HK}).
The distributions of
$(H-K)_0$
for cluster members and field stars are similar to those for the typical
star forming regions such as the Trapezium cluster \citep{Muench2002}:
the peak of distribution is at $\sim 0.2$ mag and $\sim 0.5$ mag for
field stars and cluster members, respectively.
The difference of the average 
$(H-K)_0$
of field stars and cluster members
(0.21 mag) can be attributed to thermal emission of
circumstellar disks of cluster members. Assuming that disk emission
appears in K-band and does not appear in H-band, 
the disk color excess
of the cluster members in K-band, $\Delta K_{\rm disk}$, is equal to
0.21 mag. 

\subsection{Stellar Density} \label{sec:SD}
Stellar density of the Cloud2-N cluster was estimated at $\sim 10$
pc$^{-2}$ from the 
spatial distribution of the identified cluster members
in the $\sim 2 \times 2$ pc$^2$ area.
Star forming regions show a wide range of stellar density from 
%
$1-10$
pc$^{-2}$ for T-associations to $\sim 100-1000$ pc$^{-2}$
for OB-associations 
in the area of $\sim 1-2$ pc diameter
 \citep[e.g.,][]{Lada1999}.
The estimated stellar density suggests, therefore, that
T-association-type star formation is on-going in Cloud2-N.

\subsection{K-band Luminosity Function\ (KLF)} \label{sec:KLF}
\begin{figure}[htpb]
\includegraphics[angle=-90,scale=.50]{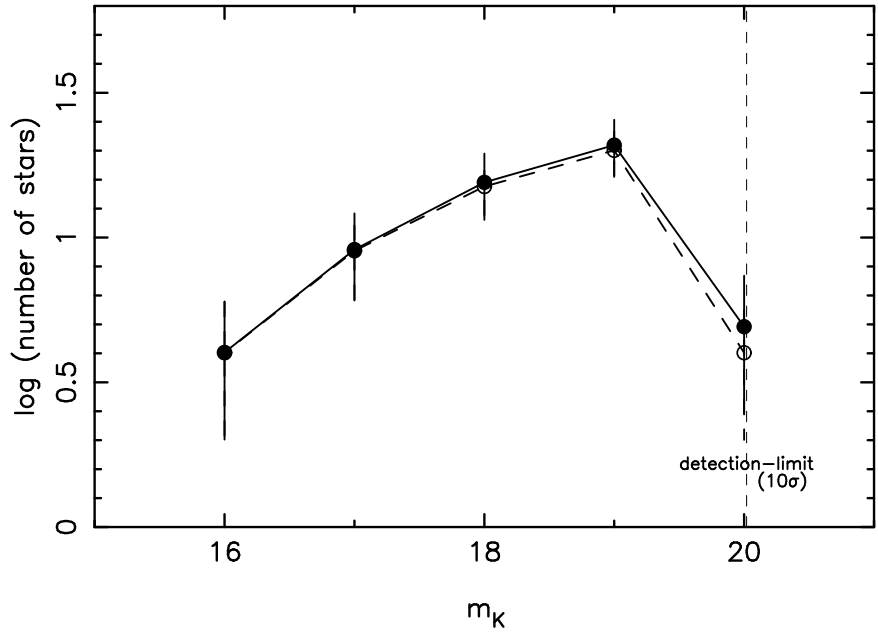}
\caption{The raw KLF and the completeness-corrected KLF of the Cloud2-N
 cluster are shown in the 1 mag bin with 
open circles
and filled circles,
respectively.
Error bars show the uncertainties due to the Poisson statistics.}
\label{fig:compKLF}
\end{figure}

We constructed a K-band luminosity function (KLF) of the identified
cloud members with the K-band apparent magnitude plotted on the
horizontal axis and the number of stars plotted on the vertical axis.
We estimated the detection completeness in each magnitude bin by putting
artificial stars on random positions in the field and checking whether
they are detected in the same way as for the photometry.  Five stars are
placed at a time in each magnitude bin and the check was conducted 200
times, resulting in about 1000 artificial stars per magnitude bin. The
result is summarized in Table~\ref{tab:comp}, and KLFs before and after
the completeness correction are shown in Fig~\ref{fig:compKLF}.

\section{Model KLF} \label{sec:modelKLF}
Although our final goal is to find out the IMF and star forming
efficiency in low-metallicity environment in EOG, many physical
parameters in such an environment are unknown. Therefore, as a first
step, we assumed 
the typical IMFs in the field and in the near-by star clusters,
and tried to see whether the resultant other parameters are consistent
or not.
The ``typical IMFs'' include the cluster IMF in the
Trapezium \citep{Muench2002}, IMFs by \citet{Miller1979},
\citet{Scalo1998}, and the average IMF by \citet{Kroupa2002}.  Among
many observed cluster IMFs, we used the Trapezium IMF because it is the
most reliable IMF for young clusters
\citep[e.g.,][]{AnnualReport}. Figure~\ref{fig:IMFs} shows all these
IMFs with their originally-defined mass ranges.

We constructed simple model KLFs for selected
ages with the following procedures:
1) assume an IMF,
2) convert the mass function to a luminosity function using a
mass-luminosity (M-L) relation from an isochrone model,
3) convert the luminosity to a K absolute magnitude, $M_K$,
with bolometric correction,
4) convert the $M_K$ to an apparent magnitude $m_K$ with the 
distant modulus of Cloud 2. 
We repeated this process for all four IMFs.
In this process we assumed that all stars are formed instantaneously
without any age spread.
We constructed model KLFs with ages of 0.1, 0.5, 1, 2 Myr.
The isochrone by \citet{{D'Antona1994},{D'Antona1998}}, and the
bolometric correction by \citet{Muench2000} were applied.
There is no isochrone model for low-metallicity environment like Cloud 2
($\sim -0.7$ dex).  However, the M-L relation for Cloud 2 is expected to
be similar to that for solar metallicity in view of the very little
difference of M-L relation for the solar metallicity and the metallicity
of $-$0.3 dex \citep{{D'Antona1994},{D'Antona1998}}.  Therefore, we
assumed that the M-L relation for solar metallicity can be applied to
the Cloud2-N cluster.
The mass range used for constructing the KLF is
 0.017 $-$ 3 M$_\odot$, which is the range of the above isochrone.

Lastly considering the $A_V$  and $\Delta K_{\rm disk }$ estimated
in section~\ref{sec:AvDISK}, the model KLF was shifted to the fainter
magnitude side by 0.72 mag ($A_K$) and to the brighter
magnitude side by 0.21 mag ($\Delta K_{\rm disk}$).
Finally the model KLFs were normalized so that numbers of stars in the
magnitude range 16.5$-$19.5 mag are the same as those for the Cloud2-N
cluster.  The resultant model KLFs are shown in Fig.~\ref{fig:KLF} with
the completeness-corrected KLF of the Cloud2-N cluster. 
The mass of the star with $m_k = 16.0$ mag is
$\sim$ 2.5M$_\odot$ and that with $m_k = 20.0$ mag is $\sim$
0.12M$_\odot$ at the age of 0.5 Myr.  
As shown in Fig.~\ref{fig:KLF},
the mass range used for constructing the KLFs (0.017 $-$ 3 M$_\odot$)
sufficiently covers the observed KLF at the fainter magnitude side while
the coverage is marginal at the brightest magnitude side especially for
older ages ($\ge 0.5$ Myr). \citet{Muench2000} pointed out that the
insufficient mass range can cause an underestimate of the number of
stars near the edge of the magnitude range (see section~3.1.3 and the
upper panel in Fig.~5 in \citet{Muench2000}).  In our case, this can
affect only the brightest magnitude bin of the model KLFs. However, the
shape of the model KLFs is well-determined without this magnitude bin
and we conclude that the following discussions, especially on age
determination, are not affected by this effect.

\section{DISCUSSION} \label{sec:DIS}

\subsection{Age of the Cloud2-N cluster} \label{sec:Age}
Because KLFs of different ages are known to have different peak magnitudes,
the age of the young clusters can be estimated. 
The comparison of the observed KLF with the model
KLFs in Figure~\ref{fig:KLF} suggests that the age of the Cloud2-N
cluster is less than 1 Myr and most likely no more than 0.5 Myr.
It is difficult to estimate the age of the cluster with an accuracy of 
0.1 Myr because isochrone models for ages of less than 1 Myr is thought
to be uncertain \citep{Baraffe}. 
However, we can at least conclude that the age of the Cloud2-N cluster
is no more than 1 Myr from the above comparison (Fig.~\ref{fig:KLF}).

There are still two uncertainties which have not been considered in the
model KLFs.
First, the distance to Cloud 2 has a moderate uncertainty around the
most likely value $D = 12$ kpc  (distance modulus (DM) = 15.4
mag) (see section~\ref{sec:INTRO}).
However, we found that the estimated age from model KLFs is still
$0.5-1.0$ Myr for the nearest possible distance ($D=8.5$ kpc, DM = 21.1
mag) and the age is even younger than 0.1 Myr for the most 
distant
case ($D=16.6$ kpc, DM = 14.6 mag). 
The second uncertainty comes from the possible large
difference of IMF from the typical IMFs because of the special
environment of the EOG. In this case, the age of the cluster could be
much older than 1 Myr. The comparison of the observed KLF and the model
KLFs for ages more than 2 Myr suggests that the IMF could have been
significantly weighted to higher mass compared to the typical IMFs if
the age of the cluster were more than 2 Myr.  
To check this we tried to
fit the slope of the observed KLF in $m_K$ = 16 $-$ 19 range assuming
the age of 2 Myr (see the dot-dashed line in Figure \ref{fig:KLF}d). The
necessary IMF slope for this fitting required a very unrealistic slope
of the IMF at the stellar mass $>$ 1 M$_\odot$ (see the grey dashed line
in Figure \ref{fig:IMFs}). This is also highly
unlikely because the stellar density of the cluster suggests that it is
T-association (see section~\ref{sec:SD}), which does 
not have a significant
number of high-mass stars. Therefore, even with the
uncertainties, we can conclude that the age of the Cloud2-N cluster is
less than 1 Myr.

%

\subsection{Comparison with Other Embedded Clusters in the Far Outer Galaxy} \label{subsec:SFEOG}


\citet{Santos2000} discovered two distant embedded young clusters in far
outer Galaxy ($R_g = 16.5$ kpc; $D = 10.2$ kpc) with near-infrared
imaging of a CO cloud associated with an IRAS source. Their
detection-limit ($m_K = 16.4$mag) corresponds to a 1 Myr $\sim 1-2
M_\odot$ star seen through 10 mag of visual extinction.  Snell,
Carpenter, \& Heyer (2002) has conducted a comprehensive NIR survey of
embedded clusters in far outer Galaxy ($R_g = 13.5-17.3$ kpc) based on
the 
FCRAO CO survey in the far outer Galaxy. They have found 11 embedded
clusters with the detection limit of $m_{K'} = 17.5$ mag.  For the most
distant cluster, this magnitude corresponds to a 1 Myr old 0.6 $M_\odot$
star with no extinction, or a 1 Myr old 1.3 $M_\odot$ star seen through
10 mag of visual extinction. These pioneering works revealed the
existence of embedded clusters in the far outer Galaxy for the first
time. 
Most importantly Snell et al.'s comprehensive work found that the star
formation activity that is similar to that found throughout the Galaxy
is ubiquitously present in the far outer Galaxy.


Our NIR imaging is the first deep NIR imaging of the embedded cluster in
the far outer Galaxy with the detection limit of $m_K \sim 20$, which is
3-4 mag deeper than the previous studies. Assuming that the age
of the Cloud2-N cluster is 0.5 Myr old (section~\ref{sec:Age}), the mass
of the brightest star ($m_K \sim 16$) is $\sim 2.5 M_\odot$ and the mass
of the faintest star ($m_K \sim 20)$ is $\sim 0.12 M_\odot$, taking into
account
A$_V$ = 6.4 mag and the disk color excess $\Delta K_{disk} = 0.21$ mag
(section~\ref{sec:AvDISK}). Despite the largest distance ($R_g=19$ kpc;
$D=12$ kpc) of the Cloud 2-N cluster among the embedded clusters in the
far outer Galaxy, our deep imaging has reached to the limit close to the
substellar mass range for the first time.  Because the model KLFs based
on the typical IMFs reasonably fit the observed KLF including the peak
near $m_k = 19$, we suggest that the IMF of the Cloud 2-N cluster can be
approximated with the typical IMFs. An assessment to more accurate IMF
requires an independent study of the age of the cluster based on
spectroscopy of each cloud members.



\citet{Snell2002} found $\sim 25 - 95$ cluster members for the 11
embedded clusters in the far outer Galaxy with the detection limit of
$M_K = 2.5 - 3.5$ mag ($m_{K'} = 17.5$ mag). \citet{Santos2000} found $\sim
30$ cluster members for a cluster in the far outer Galaxy with the
detection limit of $M_K = 1.4$ mag ($m_K = 16.4$ mag). 
The numbers of stars with those detection limits are consistent with
that in the near-by embedded clusters within 1 kpc from the solar
system. We have detected 20 cluster members in Cloud2-N down to $M_K =
2.5$ mag and 40 cluster members down to $M_K = 3.5$ mag.
Therefore, the properties of the Cloud 2-N cluster appear to be quite
similar to the star clusters found by Snell et al and Santos et al.

\subsection{Supernova Triggered Star Formation} \label{SNTSF}
\citet{NK2006}
suggested that the supernova remnant (SNR) HI shell 
\citep[GSH138-01-94;][]{Stil2001}
triggered the star formation activity in Cloud 2.
%
%
The uniform extinction of the cluster
members ($A_V \sim 6$ mag) measured in section~\ref{sec:AvDISK} is 
close to
the total extinction of Cloud2-N that is estimated at $A_V \sim 9$ mag 
from $^{13}$CO data \citep{Digel1994},
suggesting that the formation of the Cloud2-N cluster occurs behind Cloud 2. 
The geometry of the SNR HI shell, Cloud 2-N, and Cloud2-N cluster is
summarized in Fig.~\ref{fig:GEO}.
In section~\ref{sec:Age},
the age of Cloud2-N cluster was estimated at $\sim$ 1 Myr, which is 
much less than that of the SNR HI shell (4.3 Myr).
%
These
facts strongly suggest
 that the star formation
activity in Cloud 2 was triggered by the SNR HI shell.

\acknowledgments
We are grateful to August Muench for kindly providing us his KLF data
on Trapezium. 
We thank the anonymous referee for the careful reading and the positive
comments, which significantly improved our paper.
The data presented here was obtained during the
commissioning phase of the Subaru Telescope and IRCS. We truly thank
all the Subaru 
staff who made these observations possible.

\clearpage

\input{tab1}

\clearpage

\begin{figure} 
\includegraphics[angle=-90, scale=0.5]{f6.eps}
\caption{
The typical IMFs used for the model KLF
fitting. 
Red: Trapezium IMF by \citet[]{Muench2002}, Blue: gaussian-modeled IMF by
\citet{Miller1979}, Green: IMF by \citet{Scalo1998}, Orange: average IMF
by \citet{Kroupa2002}. The Kroupa IMF has two slopes at $M \ge 1
M_\odot$: Scalo slope ($\alpha = 2.7$) and Salpeter slope ($\alpha =
2.3$).  All IMFs are normalized at 1 $M_\odot$.  The dot-dashed grey
line shows an ``unrealistic'' model IMF which fits to the observed KLF
of the Cloud 2-N cluster in case the age is forced to be set at 2
Myr. See the main text in section \ref{sec:Age} for the detail.
} \label{fig:IMFs}
\end{figure}

\begin{figure}
\includegraphics[angle=-90,scale=.60]{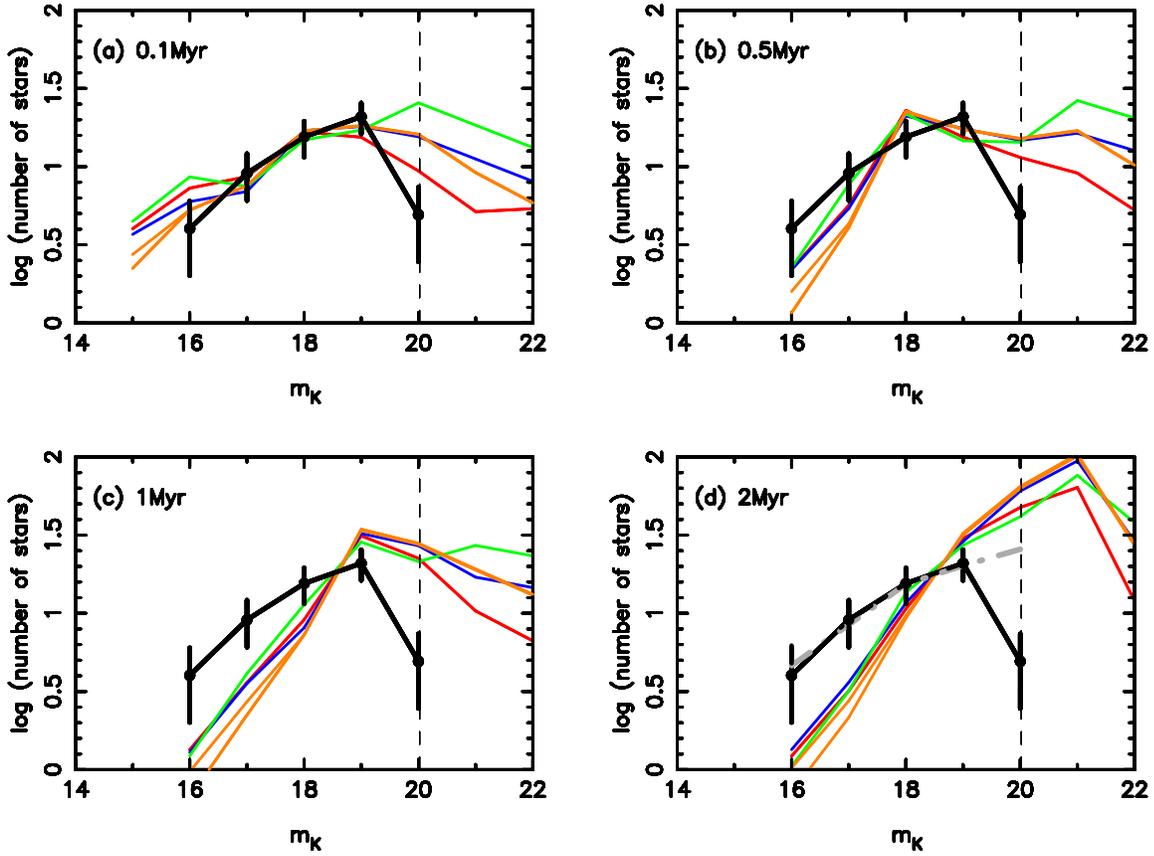}
\caption{
Comparison of completeness-corrected KLF (thick
line) and model KLFs of various ages (thin lines) for the Cloud2-N
cluster.  The underlying IMFs for the model KLFs are as shown in Figure
\ref{fig:IMFs} with the same colors.
}\label{fig:KLF}
\end{figure}


\begin{figure}
\includegraphics[scale=0.6]{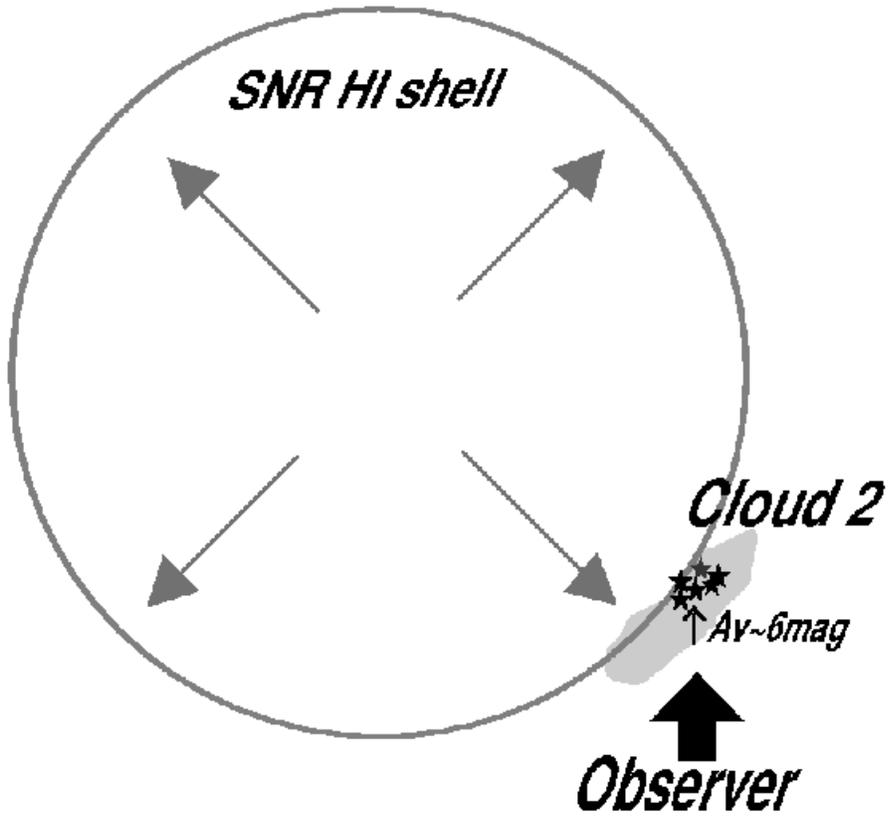}
\caption{Inferred geometry 
of the SNR
HI shell (GSH 138-01-94), Cloud 2, and Cloud2-N 
cluster. Star formation
in Cloud 2 is occurring on the shock-front 
of the SNR HI shell.}
\label{fig:GEO}
\end{figure}




%

\end{document}

%% file: tab1.tex
\ProvidesFile{table1.tex}%
 [2003/12/12 5.2/AAS markup document class]%
\begin{table}
\begin{center}
\caption{Number of detected cluster members and
 completeness. \label{tab:comp}\newline }
\begin{tabular}{cccc}
\tableline\tableline
K & completeness & $n_{\rm raw}$ & $n_{\rm cor}$ \\

(1) & (2) & (3) & (4) \\
\tableline
16.0 & 0.998 & 4  & 4.01  \\
17.0 & 0.990 & 9  & 9.09  \\
18.0 & 0.967 & 15 & 15.51 \\
19.0 & 0.959 & 20 & 20.86 \\
20.0 & 0.813 & 4  & 4.92  \\
\tableline
\end{tabular}
\end{center}
\tablecomments{Col.(1): K-band magnitude.
Col.(2): Estimated detection completeness.
Col.(3): Raw number of detected cluster members in the 1 mag bin.
Col.(4): Corrected number of cluster members in the 1 mag bin.}
\end{table}


